\documentclass[sigconf]{acmart}

\usepackage{enumitem}
\AtBeginDocument{%
  \providecommand\BibTeX{{%
    \normalfont B\kern-0.5em{\scshape i\kern-0.25em b}\kern-0.8em\TeX}}}

\usepackage{tikz}
\newcommand*\circled[1]{\tikz[baseline=(char.base)]{
            \node[shape=circle,draw,inner sep=2pt] (char) {#1};}}
\begin{document}

\title{Enhancing Data Space Semantic Interoperability through Machine Learning: a Visionary Perspective}


\definecolor{idcolor}{HTML}{A6CE39}
\newcommand{\orcidlink}[1]{\href{https://orcid.org/#1}{\color{idcolor}\faOrcid}}

\author{Zeyd Boukhers}
\affiliation{%
  \institution{Fraunhofer Institute for Applied Information Technology FIT and University Hospital Cologne, Institute for Biomedical Informatics}
  \streetaddress{ Konrad-Adenauer-Straße, 53757}
  %\city{Sankt Augustin}
  \country{Germany}
  \postcode{43017-6221}
}
\email{zeyd.boukhers@fit.fraunhofer.de}
\orcid{0000-0001-9778-9164}

\author{Christoph Lange}
\affiliation{%
  \institution{Fraunhofer Institute for Applied Information Technology FIT and RWTH Aachen University}
  \streetaddress{ Konrad-Adenauer-Straße, 53757}
  %\city{Sankt Augustin}
  \country{Germany}
  \postcode{43017-6221}
}
\email{christoph.lange-bever@fit.fraunhofer.de}
\orcid{0000-0001-9879-3827}

\author{Oya Beyan}
\affiliation{%
  \institution{University of Cologne, Faculty of Medicine and University Hospital Cologne, Institute for Biomedical Informatics}
  \streetaddress{Kerpener Straße 62 50937}
  %\city{Cologne}
  \country{Germany}}
\email{oya.beyan@uni-koeln.de}
\orcid{0000-0001-7611-3501}

\renewcommand{\shortauthors}{Trovato and Tobin, et al.}

\begin{abstract}

Our vision paper outlines a plan to improve the future of semantic interoperability in data spaces through the application of machine learning. The use of data spaces, where data is exchanged among members in a self-regulated environment, is becoming increasingly popular. However, the current manual practices of managing metadata and vocabularies in these spaces are time-consuming, prone to errors, and may not meet the needs of all stakeholders. By leveraging the power of machine learning, we believe that semantic interoperability in data spaces can be significantly improved. This involves automatically generating and updating metadata, which results in a more flexible vocabulary that can accommodate the diverse terminologies used by different sub-communities. Our vision for the future of data spaces addresses the limitations of conventional data exchange and makes data more accessible and valuable for all members of the community.

\end{abstract}

\begin{CCSXML}
<ccs2012>
   <concept>
       <concept_id>10002951.10002952.10003219.10003217</concept_id>
       <concept_desc>Information systems~Data exchange</concept_desc>
       <concept_significance>500</concept_significance>
       </concept>
   <concept>
       <concept_id>10002951.10002952.10002971.10003450</concept_id>
       <concept_desc>Information systems~Data access methods</concept_desc>
       <concept_significance>500</concept_significance>
       </concept>
   <concept>
       <concept_id>10002951.10003260.10003309.10003315</concept_id>
       <concept_desc>Information systems~Semantic web description languages</concept_desc>
       <concept_significance>500</concept_significance>
       </concept>
 </ccs2012>
\end{CCSXML}

\ccsdesc[500]{Information systems~Data exchange}
\ccsdesc[500]{Information systems~Data access methods}
\ccsdesc[500]{Information systems~Semantic web description languages}

\keywords{data spaces, semantic interoperability, machine learning}

\iffalse
\begin{teaserfigure}
  \includegraphics[width=\textwidth]{sampleteaser}
  \caption{Seattle Mariners at Spring Training, 2010.}
  \Description{Enjoying the baseball game from the third-base
  seats. Ichiro Suzuki preparing to bat.}
  \label{fig:teaser}
\end{teaserfigure}
\fi
\received{20 February 2007}
\received[revised]{12 March 2009}
\received[accepted]{5 June 2009}

\maketitle

\section{Introduction}

Data
spaces are a cutting-edge solution to the difficulties encountered in data exchange and integration. They act as a federated platform for sharing and exchanging data among various entities, providing the necessary tools and security measures to ensure data can be safely shared and consumed. The ultimate aim of data spaces is to enhance the accessibility and usage of data by a wide range of stakeholders, fostering innovation and research in the digital economy. These innovative infrastructures empower all users by making data integration simpler, more adaptable, and more efficient. conventional data exchange systems, data spaces have the advantage of easily onboarding new members.  

%Unlike conventional data exchange systems, dataspaces offer an incremental approach to data integration, allowing for a gradual implementation of advanced features as the need arises, starting with basic functionality.

The theoretical importance of data spaces lies in their ability to seamlessly integrate data sources that have diverse schemas and structures, without the requirement of a shared schema or extensive upfront effort to standardize the data. This is made possible through the extensive use of semantic technologies~\cite{alexiev2022data}, which help align the data and make it usable in a consistent manner. As a result, using data spaces reduces the time and effort needed to establish a data integration system, while offering increased flexibility and scalability. However, achieving satisfactory levels of semantic interoperability in data spaces is a complex and ongoing process. It requires ongoing development of standards, tools, and techniques for data integration and processing. %Moreover, the need for privacy and security in data spaces can pose additional challenges to achieving semantic interoperability, as entities may be reluctant to share their data if they are concerned about data privacy and security.

International Data Spaces\footnote{\url{https://internationaldataspaces.org/}} and Gaia-X\footnote{\url{https://gaia-x.eu/}} are two prominent initiatives in Europe and globally that are designed to bring together various stakeholders to create a secure, interoperable, and decentralized data infrastructure. The ultimate objective of IDS and Gaia-X is to establish a trustworthy, secure, and efficient data infrastructure that supports data-driven innovation while preserving privacy and security. These initiatives focus on providing infrastructure services that increase trust among entities, advance data sovereignty, and protect data privacy. As a result, they are expected to simplify the process of accessing and utilizing data securely and transparently for businesses and organizations.

As the number of organizations and individuals participating in various data spaces continues to increase, the demand for effective data management and exchange solutions will also rise. The key to successful data exchange lies in the mutual understanding of the data being shared, which is referred to as \emph{semantic interoperability}. Thus, finding ways to enhance semantic interoperability is of utmost importance as the number of members joining different data spaces continues to grow.

Although machine learning has shown promise in addressing the above-mentioned interoperability aspects of data exchange in general, its application to Data Spaces has not been fully explored. This paper aims to shed light on the potential of using machine learning to improve semantic interoperability in Data Spaces, making it the go-to solution for data exchange.

In this paper, we present our perspective on how machine learning solutions can be utilized to improve the semantic interoperability of a data space, using IDS as a concrete example. Although there are numerous aspects of semantic interoperability, we concentrate on six key challenges that are prevalent in data exchange. It is crucial to note that this paper does not make any definitive claims, but instead offers a comprehensive overview and framework for integrating machine learning solutions directly into data spaces.

\section{Related Work}

%Data Spaces as a research project has started in 2015. Therefore, there the focus was on the legal, technical and metadata interoperability aspects and only a few studies have addressed the data semantic aspect~\cite{alexiev2022data}. 

The concept of Data Spaces has been in existence for several decades, but in recent times, it has gained significant attention, and considerable effort is being devoted to facilitating data exchange in today's data-driven ecosystem, such as International Data Spaces (IDS)\footnote{\url{https://www.fraunhofer.de/en/research/lighthouse-projects-fraunhofer-initiatives/international-data-spaces.html}} and the Common European Data Spaces~\cite{scerri2022common}. 
So far, the primary focus in practical data spaces has been on legal, technical, and metadata interoperability, with little attention given to the semantic aspect of data, as only a few studies have been conducted in this area~\cite{alexiev2022data}. This means that in terms of semantic interoperability, the current focus is on metadata only with the assumption that it exists. However, data semantic interoperability has been studied for decades. For example, Ouksel et al.~\cite{ouksel1999semantic} discussed the issue of finding accurate information in a complex, heterogeneous information system like the Internet and Web. They proposed a framework for interoperability that involves relating information to real-world entities and acknowledges the changing nature of semantics. More than one decade later, Kiljander et al.~\cite{kiljander2014semantic} discussed the need for common approaches to enable high-level interoperability between heterogeneous IoT devices to realize pervasive computing and IoT visions. It divides the interoperability challenge into two levels: connectivity and semantics. The connectivity level covers traditional Open System Interconnection (OSI) model layers from the physical to the transport layer. The semantic level covers technologies needed for enabling meaning-sharing between communicating parties. The authors stated that semantic level interoperability has been identified as a main goal in the Semantic Web and that semantic web technology can be used to represent knowledge about the physical world in IoT-related projects.

%EU Data Spaces

Semantic interoperability in data exchange has been also addressed in specific domains. Lin et al.~\cite{lin2011investigating} evaluated the usage of Logical Observation Identifiers Names and Codes (LOINC) and its impact on the interoperability of laboratory data from different institutions that use LOINC codes. Heterogeneous data formats have been discovered among different institutions for the same laboratory tests using LOINC codes. After investigating the common problems that arise when aggregating such data, they suggest that more guidance on best practices in coding laboratory results is needed to achieve greater interoperability.

\section{ML-enhanced Data Spaces}

\begin{figure}
    \centering
    \includegraphics[width=\linewidth]{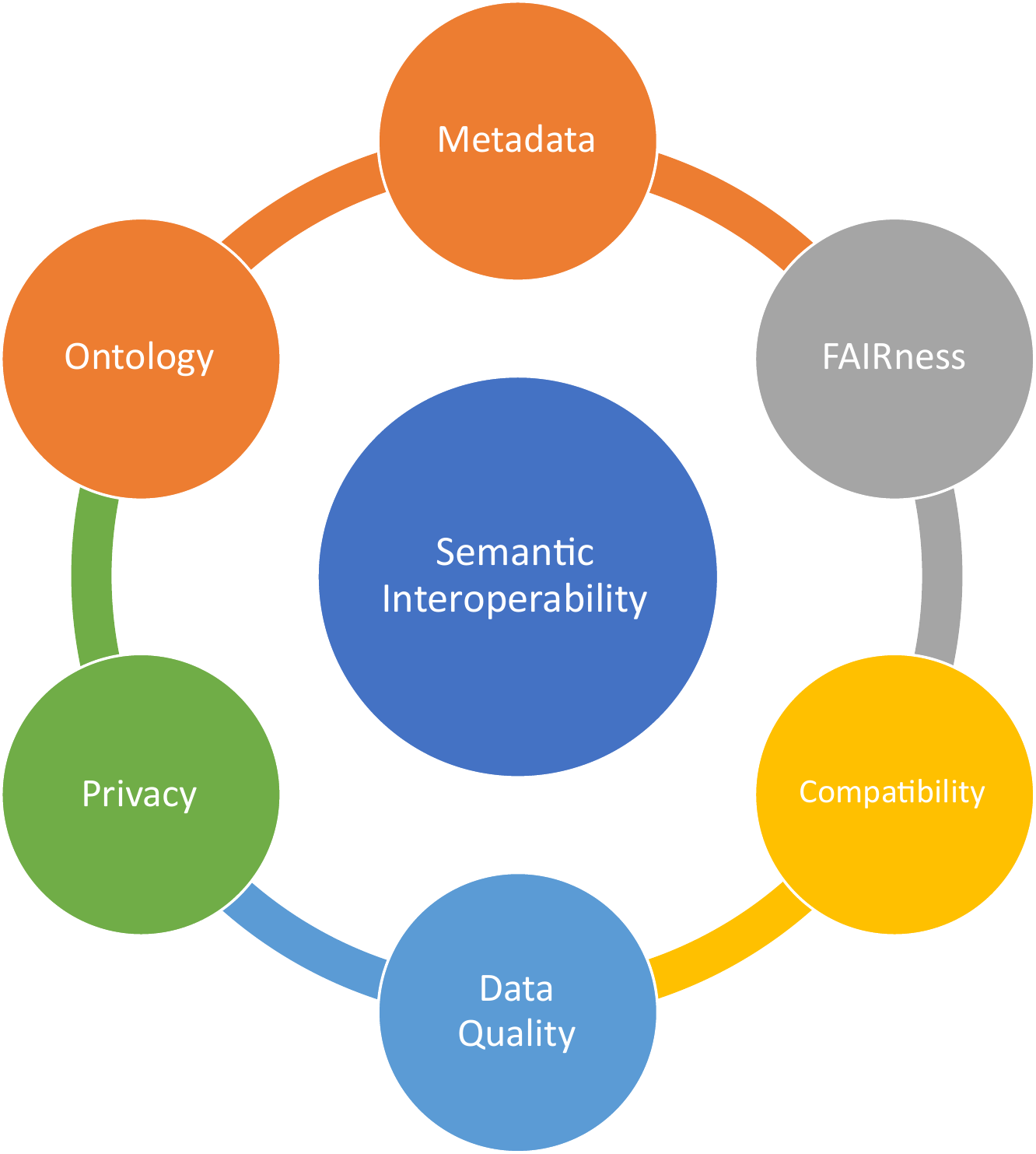}
    \caption{Semantic interoperability aspects in data spaces that machine learning can enhance}
    \label{fig:aspects}
\end{figure}

Semantic interoperability in data spaces is a complex issue that involves multiple aspects, as illustrated in Figure~\ref{fig:aspects}. While machine learning has the potential to improve each of these aspects, traditional approaches have primarily utilized machine learning techniques in isolation, rather than within the broader context of data spaces. It is vital to consider the full spectrum of semantic interoperability aspects and integrate machine learning in a comprehensive and holistic manner within the data space environment. 

\begin{figure*}[h]
    \centering
    \includegraphics[width=\linewidth]{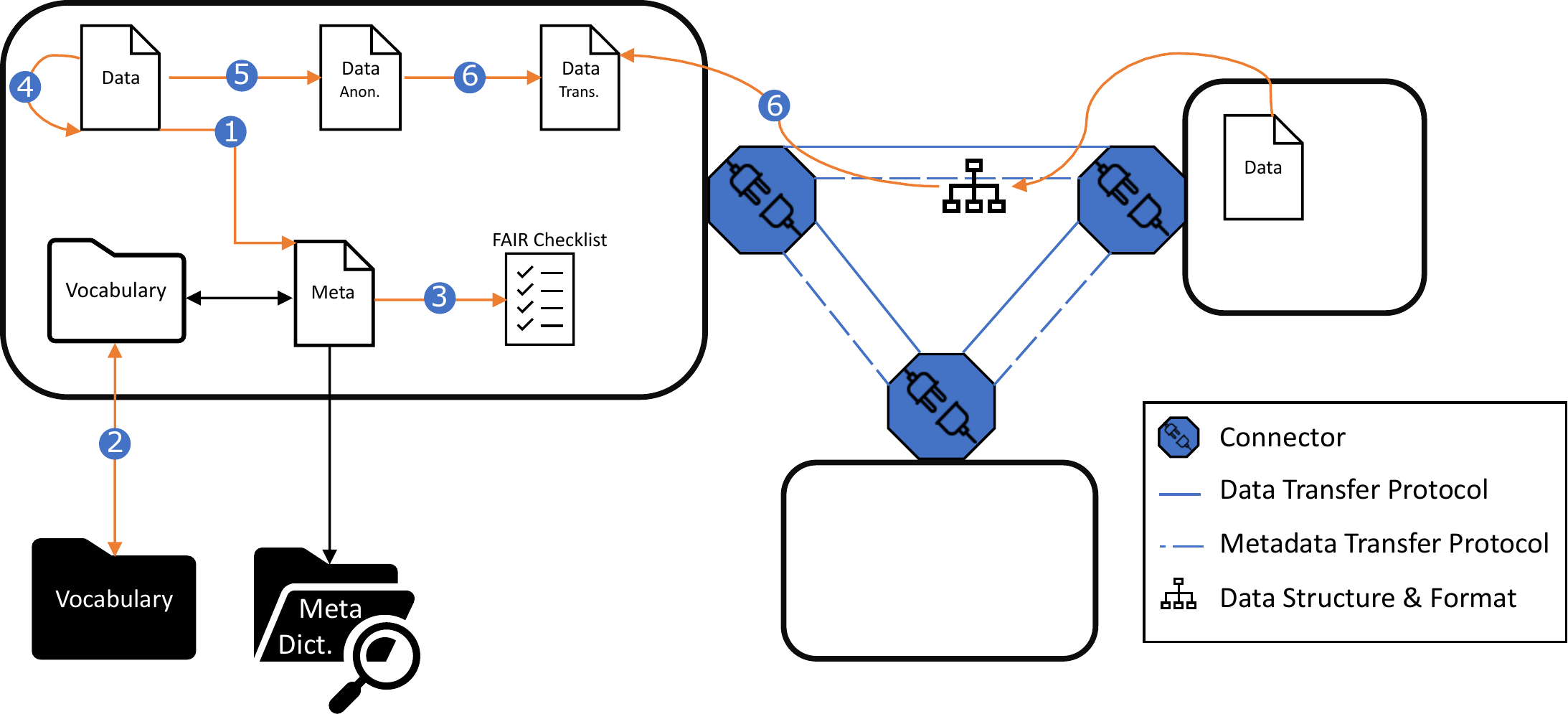}
    \caption{An overview of an ML-enhanced Data Space with three members. (1) Automatic Metadata Extraction, (2) Ontology and Vocabulary Alignment, (3) FAIRness Evaluation, (4) Data Quality Assessment \& Enhancement, (5) Privacy Preserving, (6) Compatibility Improvement.}
    \label{fig:overview}
\end{figure*}

Figure~\ref{fig:overview} presents an overview of the ML-enhanced data space in the International Data Spaces environment, showcasing six key aspects of data management among three stakeholders, including data providers and consumers and service providers. These aspects are:
\begin{itemize}[leftmargin=*]
\item  \textbf{Automatic Metadata Extraction (\circled{1})}: A machine learning model can automatically extract essential attribute values from the data if metadata is not already available, helping data providers to prepare their data for exchange and consumption without the need for manual metadata preparation.

\item \textbf{Ontology and Vocabulary Alignment (\circled{2})}: The vocabulary of the data space is aligned with the vocabulary of the data provider, enabling data consumers to understand the data being exchanged. This eliminates the need for members in the data space to adopt the same internal vocabulary, which can often be a challenging task.

\item \textbf{FAIRness Evaluation (\circled{3})}: The FAIRness level of the data is assessed based on provided or extracted metadata, allowing the data provider to improve the FAIRness of their data and allowing the data consumer to understand the ease of use of the data.

\item \textbf{Data Quality Assessment \& Enhancement (\circled{4})}: The quality of the data is evaluated and improved if possible, based on the format of the data. Machine learning can be used to evaluate and enhance structured and tabular data, however, it's important to recognize that the quality metrics may vary depending on the format of the data. For example, it might be challenging to assess the quality of unstructured data (e.g., a corpus of documents).

%not all formats support quality assessment.

\item \textbf{Privacy Preserving (\circled{5})}: ML-based anonymization and masking techniques can be applied to data that contains private, sensitive, or personal information to make it shareable. Sensitive data can be automatically detected or provided by the data provider, allowing data providers to share their data without any privacy concerns.

\item \textbf{Compatibility Improvement (\circled{6})}: The data is transformed into a readable format for the data consumer. In cases where data is being merged with the consumer's data, the consumer will communicate the structure and format, enabling the data to be transformed accordingly. This allows the consumer to make use of the received data without having to put in additional effort to read and understand it.
\end{itemize}

In the following, we discuss each of these aspects:

\subsection{Automatic Metadata Extraction}
Metadata plays a vital role in data exchange as it enables data consumers to understand the data and determine if it meets their needs. However, many data providers may be hesitant to provide the necessary metadata due to a lack of capacity or knowledge to prepare it for their resources. This can be a significant obstacle in data exchange, as it limits the ability of consumers to access and utilize the data they need.

To overcome this challenge, machine learning can be leveraged to (semi) automatically extract metadata from resources. Machine learning algorithms can be trained on a dataset of resources and their corresponding metadata, allowing them to learn the patterns and relationships between the data and the metadata. These algorithms can then be applied to new resources to extract the relevant metadata. This approach has the advantage of being able to handle complex and nuanced relationships between the data and the metadata. It can also be easily updated and adapted as the data and its needs evolve. However, it is important to note that a typical challenge in data spaces is that the resources have different, heterogeneous formats.

%Machine learning can also assist in the creation of metadata for resources by using natural language processing to extract information from the resource's title or description to automatically generate the metadata. The metadata properties can vary based on the resource being exchanged and it's necessary to use different ML models for different sources.

Different resources being exchanged in data spaces can have varying metadata properties, and it may be necessary to utilize different machine learning (ML) models for different resources and metadata attributes. For instance, in the case of document corpora, Natural Language Processing (NLP) techniques can be employed to extract titles and descriptions. Specifically, automatic metadata extraction techniques such as those in~\cite{boukhers2022vision, tkaczyk2017new} can be utilized to extract metadata from each document, such as \emph{Publication Date}, \emph{Author}, \emph{Language}, etc. This metadata can then be used to derive the metadata for the entire collection, such as \emph{Publication Range}, \emph{Authors}, \emph{Languages}, etc.

\subsection{Ontology and Vocabulary Alignment}
\label{sec:onto}

The International Data Spaces Reference Architecture\footnote{\url{https://internationaldataspaces.org/use/reference-architecture/}} highlights the importance of common vocabularies for effective data exchange within a data space. However, in practice, data providers may have their own unique vocabularies, making it difficult to align them with the vocabulary used in the data space. This can be due to the cost and effort involved in mapping their existing vocabularies to the data space vocabulary, or due to the fact that a data provider may participate in multiple data spaces with different vocabularies.

To tackle these challenges, machine learning algorithms can be utilized to support automatic mapping between the local vocabulary of a data provider and the vocabulary used in the data space. This allows for seamless and interoperable data exchange, without requiring data providers to adopt a new vocabulary.

Machine learning-based methods for ontology alignment~\cite{nezhadi2011ontology} and ontology matching~\cite{doan2004ontology} can be applied to automatically map concepts and terms from one ontology or vocabulary to another. These algorithms use techniques such as semantic similarity measures~\cite{sousa2022supervised}, graph-based methods~\cite{shenoy2013secured}, and deep learning models~\cite{khoudja2018ontology, iyer2020veealign, bento2020ontology} to identify correspondences between concepts in different ontologies or vocabularies. The goal is to produce a mapping that enables data exchange between systems using different ontologies or vocabularies while preserving the meaning of the data.

\subsection{FAIRness Evaluation}

%Recently, the FAIR principles (i.e. \textbf{F}indable, \textbf{A}ccessible, \textbf{I}nteroperable and \textbf{R}esuable) are highly regarded in data exchange as compliance with them increases the likelihood of resources being reused. Therefore, evaluating the FAIRness of a resource can assess its fitness for use. For instance, even though the data might be suitable for a particular use case, the accompanying license might not be appropriate. Identifying this in advance can save time and resources, as the cost of negotiating for data exchange or waiting for access to data that does not meet all the requirements (including FAIRness, licensing, and access conditions) can take several months. Thus, evaluating the FAIRness level of a resource beforehand can help reduce the time and effort spent on obtaining data that may not be suitable for the intended use.

The FAIR (i.e., \textbf{F}indable, \textbf{A}ccessible, \textbf{I}nteroperable and \textbf{R}eusable) principles are becoming increasingly important in data exchange and sharing. These principles aim to ensure that data resources are easily discoverable, accessible, can be easily integrated with other data sources, and can be reused for multiple purposes. Compliance with these principles makes it more likely that data will be used and reused, as it increases the overall quality and usability of the resource.

Evaluating the FAIRness of a resource is a crucial step in determining its fitness for use, as it helps to identify any potential barriers to reuse. This can include issues such as licensing restrictions, data access conditions, and data interoperability issues. Conducting this evaluation in advance can save valuable time and resources, as it helps to avoid the need for costly negotiations or lengthy wait times for access to data that may not be suitable for the intended use.

As discussed in Section~\ref{sec:onto}, the use of shared vocabularies, such as ontologies, is important for increasing the findability and interoperability of resources. However, only using mapping techniques (see Section~\ref{sec:onto}) may not be enough, as internal ontologies that describe the metadata may not be represented using common classes. To address this issue, machine learning techniques, such as BERTmap~\cite{he2022bertmap}, can be used to assess the level of compatibility between the provider's ontology and the data space's ontology. Additionally, rule-based and semantic web technologies can be used to evaluate the structure of the metadata, further increasing the overall FAIRness of the resource.

\subsection{Data Quality Assessment \& Enhancement}
Data quality is a crucial concern for data consumers, as it impacts the trustworthiness and usefulness of the data. Unfortunately, metadata alone cannot provide any indication of the quality of the data. To ensure the quality of data, various dimensions must be considered, including accuracy, completeness, correctness, validity, integrity, and uniqueness. The importance of each dimension may vary depending on the intended use of the data and the needs of the data consumer.

Accuracy refers to how closely the data reflects the real-world phenomenon it represents. Completeness refers to the extent to which all necessary data is present. Correctness pertains to the degree to which the data adheres to established rules, such as those related to syntax, semantics, or data constraints. Validity refers to the degree to which the data follows the predefined format, structure, and domain. Integrity is the degree to which the data is protected against unauthorized changes. Lastly, uniqueness refers to the degree to which each data item is distinct and identifiable.

To ensure data quality, data providers must take steps to assess and improve the quality of their data. This can include implementing data validation and quality checks, using techniques like data profiling and data cleaning, and implementing data governance policies and procedures. Data consumers should also take steps to assess the quality of the data they receive, such as evaluating the data's source and provenance, performing data quality checks, and monitoring the data for anomalies.

Machine learning algorithms can play a crucial role in ensuring the quality and accuracy of data. One way they achieve this is by comparing the data to other sources to validate its accuracy. Additionally, machine learning algorithms can be trained to identify patterns and anomalies in the data~\cite{agrawal2015survey, pang2021deep}, helping to flag any potential inaccuracies or errors.

Another benefit of using machine learning algorithms is the ability to complete missing data. By analyzing patterns and relationships in the data, machine learning models can make predictions about missing values and fill them in~\cite{thomas2021systematic, raja2020missing, hasan2021missing}. This is especially useful in cases where it would be time-consuming or challenging to manually fill in missing data.

Furthermore, machine learning techniques can also be applied to identify and remove duplicates in data, improving the overall uniqueness and consistency of the data~\cite{park2022deepsketch,tarun2021scheme, christen2019towards}.

\subsection{Privacy Preserving}

Private and sensitive data, such as personal information, medical records, and financial data, is often subject to strict regulations and guidelines for protection and access.
% These regulations.
In order for different systems to exchange and use private data, they must be able to accurately interpret and understand the meaning and context of the data, and ensure that it is being used in compliance with applicable laws and regulations. ensuring semantic interoperability for private data requires a combination of technical solutions, such as secure data exchange protocols and data anonymization techniques, and strict governance and compliance mechanisms.

To achieve this, data providers can use machine learning techniques to automatically detect private and sensitive data in their systems~\cite{ray2021sensitive, ahmed2021automated} and take appropriate actions to mask~\cite{torra2022privacy} or anonymize~\cite{majeed2020anonymization} the data. This can help protect individuals' privacy while enabling data sharing and interoperability. For example, techniques such as data de-identification, data masking, and differential privacy can be used to remove identifying information from data while preserving its usefulness for analysis.

\subsection{Compatibility Improvement}
Also, when the same vocabulary and ontology are used by the data provider and consumer, resources are not semantically interoperable if they are not compatible with the consumer system of their resource to be integrated with. To overcome the incompatibility of resources in data exchange, solutions include data mapping and data transformation. Machine learning techniques have shown great performance in these tasks.

Resources are not semantically interoperable when they cannot be understood or used by the systems that need to access them. This can occur when the resources have different data formats or structures, making it difficult for systems to integrate and make use of the information.

To overcome the incompatibility of resources in data exchange, solutions include data mapping~\cite{li2018mfecnn} and data transformation~\cite{sajid2019predictive}. Data mapping is the process of aligning the data elements from one resource to the corresponding elements in another resource. Data transformation is the process of converting data from one format or structure to another. Both of these solutions can help to make resources compatible and enable data exchange. Machine learning can also be used to convert data from one format to another, such as natural language text to structured data~\cite{verma2020unstructured}.

\subsection*{Discussion}

%As the number of organizations and individuals participating in various Data Spaces continues to increase, the demand for effective data management and exchange solutions will also rise. The key to successful data exchange lies in the mutual understanding of the data being shared, which is referred to as semantic interoperability. Thus, finding ways to enhance semantic interoperability is of utmost importance as the number of members joining different Data Spaces continues to grow.

%Although machine learning has shown promise in addressing the above-mentioned interoperability aspects of data exchange in general, its application to Data Spaces has not been fully explored. This paper aims to shed light on the potential of using machine learning to improve semantic interoperability in Data Spaces, making it the go-to solution for data exchange.

The enhancement of semantic interoperability of data spaces is a complex task that involves different facets and approaches. In this paper, we have focused on specific aspects that can be improved through the use of machine learning in the context of International Data Spaces. To achieve this, we propose the development of machine learning-powered software that can be easily integrated into the Data Spaces connectors as smart data apps. This will make the software more user-friendly and accessible, allowing for seamless integration into the existing system.

In addition, with the growing popularity of Gaia-X in Europe and beyond, this software can also be provided as a service within the Gaia-X framework, offering members a valuable resource for improving semantic interoperability. By integrating machine learning into the data spaces, organizations can ensure that their data is properly structured, and their systems can effectively communicate and exchange information with other systems, resulting in more efficient and effective data management and exchange.

 \section{Conclusion}

 In this paper, we presented our innovative perspective on enhancing semantic interoperability in data spaces through the use of machine learning. Our focus was on six crucial aspects of interoperability within the International Data Spaces architecture, and we highlighted the significance of each of these aspects and how machine learning can improve their impact on successful data exchange.

As a follow-up to this work, we plan to test some of the concepts and solutions presented in this paper by integrating them into real-world data exchange scenarios in both the International Data Spaces and Gaia-X architectures. This will provide valuable insights into the practical implementation and effectiveness of our proposed approach, and help to further advance the state of the art in data interoperability and exchange.

\bibliographystyle{ACM-Reference-Format}
\bibliography{sample-base}

\end{document}